\documentclass[twocolumn,superscriptaddress,floatfix,
longbibliography,aps,pra,longbibliography,preprintnumbers,10pt]{revtex4-2}

\usepackage[utf8]{inputenc}
\usepackage[english]{babel}
\usepackage{color}
\usepackage{graphicx}
\usepackage{bm}
\usepackage{xcolor}
\usepackage{epstopdf}
\usepackage{amsmath}
\usepackage{amssymb}
\usepackage{epstopdf}
\usepackage[normalem]{ulem}
\usepackage{enumerate}
\usepackage{subfigure}

\newsavebox{\measurebox}

\usepackage[urlcolor=blue,colorlinks=true,citecolor=blue,linkcolor=blue,pdfstartview={FitH},bookmarks=false]{hyperref}
\graphicspath{{figures/}{./figures/}{.}}

\newcommand{\Imag}{\text{Im}}

\newcommand{\im}{\textrm{i}}

\newcommand{\bsig}{\bm{\sigma}}
\newcommand{\btau}{\bm{\tau}}

\newcommand{\bs}{\boldsymbol}
\newcommand{\bpm}{\begin{pmatrix}}
\newcommand{\epm}{\end{pmatrix}}

\newcommand{\blue}[1]{{\color{blue}#1}}

\DeclareMathOperator{\Tr}{tr}

\DeclareMathOperator{\sgn}{sgn}

\DeclareMathOperator{\CuL}{\underline{\mathcal{L}}}

\begin{document}

\title{Analytical and semianalytical tools to determine the topological character of Shiba chains}

\author{Nicholas Sedlmayr}
\email[e-mail: ]{sedlmayr@umcs.pl}
\affiliation{Institute of Physics, Maria Curie-Sk\l{}odowska University,
Plac Marii Sk\l{}odowskiej-Curie 1, PL-20031 Lublin, Poland}
\author{Vardan Kaladzhyan}
\affiliation{Department of Physics, University of Basel, Klingelbergstrasse 82, CH-4056 Basel, Switzerland}
\author{Cristina Bena}
\affiliation{Institut de Physique Th\'eorique, Universit\'e Paris Saclay, CEA
CNRS, Orme des Merisiers, 91190 Gif-sur-Yvette Cedex, France}

\date{\today}

\begin{abstract}
We introduce three new analytical and semi-analytical tools that allow one to determine the topological character of impurity Shiba chains. The analytical methods are based on calculating the effective Green's function of an infinite embedded chain using the T-matrix formalism and describing the chain as a {\it line impurity}. We thus provide a solution to the longstanding size-effects problem affecting the only general alternative method, the numerical tight-binding analysis. As an example we consider a chain of magnetic impurities deposited on an s-wave superconducting substrate with Rashba spin-orbit and we calculate its topological phase diagram as a function of  the magnetic impurity strength and the chemical potential. We find a perfect agreement between all our new techniques and a numerical analysis. 
\end{abstract}

\maketitle

\section{Introduction}
Magnetic impurities in conventional superconductors give rise to impurity-bound subgap states, the so-called Yu-Shiba-Rusinov states \cite{Yu1965,Shiba1968,Rusinov1969,Yazdani1997,Menard2015}.  The topological properties of chains of such magnetic impurities, also known as Shiba chains, have come into focus in recent years. Since the pioneering theoretical proposals~\cite{Nadj-Perge2013,Pientka2013,Braunecker2013,Klinovaja2013,Vazifeh2013} these chains became one of the most prominent platforms for Majorana bound states. Various alternatives have been considered theoretically~\cite{Pientka2014,Poyhonen2014,Reis2014,Kim2014a,Li2014,Heimes2014,Brydon2015,
Weststrom2015,Peng2015,Hui2015,Rontynen2015,Braunecker2015,Poyhonen2016,Zhang2016a,
Li2016c,Rontynen2016,Hoffman2016,Li2016b,Schecter2016,Christensen2016,Kaladzhyan2017b,
Andolina2017,Kobialka2020b}, while experimental work has mostly focused on magnetic islands \cite{Menard2017,Menard2019a} and on emulating magnetic Shiba chains on top of superconducting substrates \cite{Nadj-Perge2014,Pawlak2015,Ruby2015,Feldman2017a,Ruby2017,Kim2018b,Pawlak2019}. In the latter experiments, zero-bias peaks observed in the tunneling spectra at the ends of magnetic-atom chains are assumed to correspond to Majorana bound states. The true nature of these states remains, however, open to question, since the presence of zero-bias peaks is not an unambiguous proof of the topological character of the underlying system. 

Previously, the only general versatile tool to determine the topological character of an impurity chain was a numerical tight-binding analysis. Other theoretical approaches proposed were not so generally applicable: among the few alternative proposals one can mention describing the fine-tuned case of dilute impurities corresponding to Shiba states forming very close to zero energy~\cite{Pientka2013}, and approaches based either on Pfaffian invariants for effective one-dimensional models~\cite{Pientka2014,Nadj-Perge2014,Heimes2015,Kim2018b,Kobialka2020b,Weststrom2015,Schecter2016} or on the zeros of the Green's function in the presence of impurities~\cite{Slager2015a}. Other more recent techniques include for example a multiple point-impurity T-matrix \cite{Mashkoori2020}. However, the tight-binding method, when applied to a large but finite two-dimensional system with a modified charge/spin chemical potential along a finite chain of atoms, exhibits size-effects problems, and thus the computing times and resources necessary to model the chain accurately are very large. This is due especially to the interactions between the states forming at the ends of the chain, as well as due to the leakage of the latter into the two-dimensional bulk \cite{Stanescu2011a,Peng2015}.

\begin{figure}
	\centering
	\includegraphics[width=0.95\columnwidth]{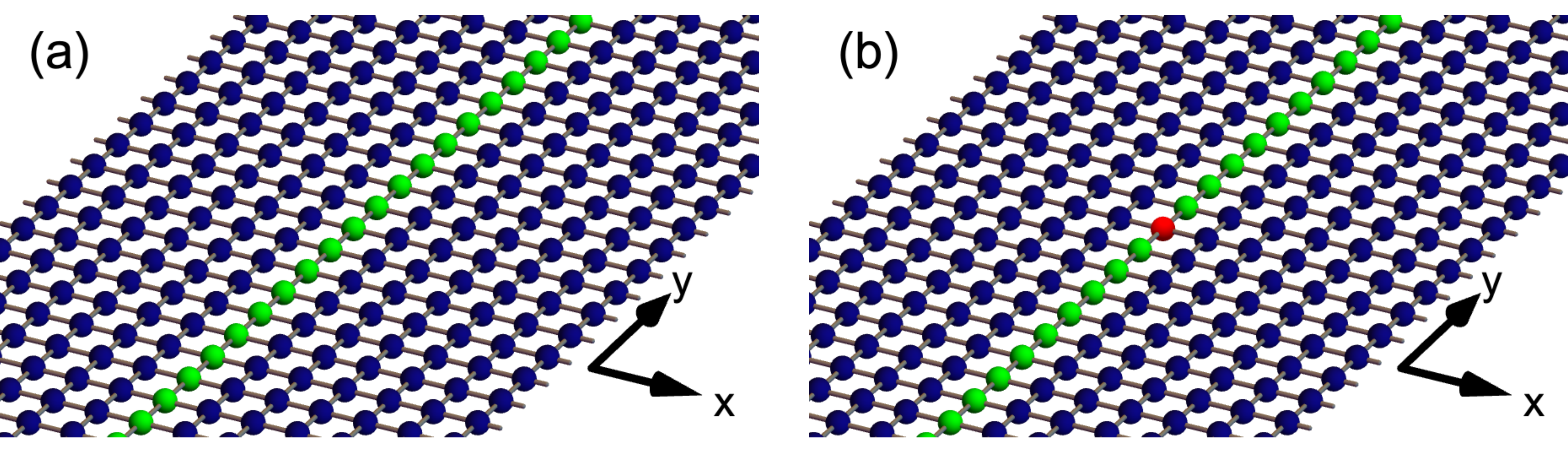}
	\caption{Schematics of the systems considered. Panel (a) shows a chain of magnetic impurity atoms (green) on the surface of the superconductor (dark blue). For I (see main text) the system is infinite in both directions. For III the surface is finite in the $x$-direction, and infinite in the $y$-direction. In panel (b) the red atom is an additional probe point-impurity, as described in II in the main text. All numerical tight-binding solutions are performed for systems finite in both directions.}
	\label{fig:systemsketch}
\end{figure}

In what follows we introduce two analytical tools and one semi-analytical tool that allow one to bypass the size-effects problem and provide us with at least one fully general and size-independent approach to test if a given impurity chain is topological. To exemplify our method we focus on a two-dimensional $s$-wave superconducting substrate with Rashba spin-orbit interaction and a chain of magnetic impurities deposited on top. 

	I.~We consider an infinite two-dimensional bulk and we model the Shiba chain as an infinite impurity line described by a finite magnetic potential along the central line of atoms (see Fig.~\ref{fig:systemsketch}).  We solve this system exactly using the $T$-matrix formalism \cite{Byers1993,Salkola1996,Ziegler1996,Mahan2000,Balatsky2006,Zazunov2016,Bena2016,Alvarado2020}. We thus obtain the effective Green's function for the impurity wire. Finally, using the latter we calculate the standard topological chiral invariant \cite{Gurarie2011} enabling us to predict in which parameter range the Shiba chain is topologically non-trivial.

	II.~In the second approach we use the effective Green's function obtained within the first approach, but we test the formation of Majorana bound states by adding an extra probe point-impurity on the wire, see Fig.~\ref{fig:systemsketch}(b). We model the impurity as an infinite-amplitude scalar potential that effectively cuts the wire in two. We check the formation of Majorana states by calculating the average zero-energy density of states in the wire in the presence of the probe impurity. If such states form, the zero-energy density of states in the wire is finite, while if the system is nontopological it is equal to zero. Additionally, one needs to check that no zero-energy states exist in the wire in the absence of the probe impurity.

 	III.~ We compare these methods with a semi-analytical approach based on Ref.~\cite{Sedlmayr2016} for calculating the $\mathbb{Z}_2$ invariant of quasi-1D systems. We consider a quasi-one-dimensional system, infinite in one direction and finite in the other. The Shiba chain is modeled the same way as in I, see Fig.~\ref{fig:systemsketch}(a) but now our system is finite in the $x$ direction. We consider systems of different width; for sufficiently wide systems the invariant no longer depends explicitly on the width.
	
The first two approaches are fully analytical and are bereft of finite-size problems, rendering the problem much more tractable and accurate than tight-binding simulations. The effective Green's function for the wire can be obtained using at most a numerical integral in momentum space, and in some cases, such as the one presented here, a closed-form expression can be derived without resorting to numerical integration.

The third approach is semi-analytical in the sense that we have a fully analytical formula for the topological invariant, but its complexity depends on the transverse size of the system, and thus for large widths it can be evaluated only numerically.

The main advantage of our approaches is that they do not depend on the size of the system, thus providing reliable tools that are not perturbed by finite-size effects. Furthermore, while the first and third approaches require some basic symmetry analysis in order to be implemented, the second one is fully general and can be effortlessly implemented and straightforwardly applied to any system.

We also want to point out that our T-matrix approach, although superficially similar to previous work, is not at all the same. Previous approaches start with a {\it point impurity} and calculate the subgap bound states for it using the T-matrix formalism,  then consider the interactions between these impurity states to construct a 1D effective Hamiltonian ~\cite{Pientka2014,Nadj-Perge2014,Heimes2015,Kim2018b,Kobialka2020b,Weststrom2015,Schecter2016}. Some also perform a multiple point-impurity T-matrix calculation which takes into account a finite number of impurities \cite{Mashkoori2020}.

In contrast, we do not start by calculating the subgap states of a {\it point impurity}, or of a system with a finite number of point impurities, but we apply the T-matrix for an infinite superconducting substrate to which we add a {\it line impurity}, corresponding to an infinite Shiba chain, for which we calculate the effective Green's function.


\section{Model} To demonstrate these approaches we focus on a two-dimensional square lattice with Rashba spin-orbit interaction and $s$-wave superconductivity, which can be described in real space by the following tight-binding Hamiltonian
\begin{align}
\mathcal{H}^{\rm 2D} =& \sum\limits_{\bs{r}} \Psi_{\bs{r}}^\dag \left( - \mu \btau^z - \Delta \btau^x \right) \Psi_{\bs{r}} \nonumber \\&
- \Psi_{\bs{r}}^\dag \left(t+\im\lambda \bsig^y \right) \btau^z \Psi_{\bs{r}+\bs{x}} + \mathrm{H.c.} \nonumber \\ &
- \Psi_{\bs{r}}^\dag \left(t-\im\lambda \bsig^x \right) \btau^z  \Psi_{\bs{r}+\bs{y}} + \mathrm{H.c.}\,,
\label{eq:H2DSC}
\end{align}
written in the basis $\Psi_{\bs{r}}=(c_{\bs{r},\uparrow},c_{\bs{r},\downarrow},c_{\bs{r},\downarrow}^\dagger,-c_{\bs{r},\uparrow}^\dagger)^\mathrm{T}$, where $c_{\bs{r},\sigma}^{(\dagger)}$ annihilates (creates) an electron with spin $\sigma = \uparrow, \downarrow$ at site $\bs{r} = (x, y)$. We denote by $\mu$ the chemical potential, $\Delta$ is the superconducting pairing parameter, $t$ the hopping constant, and $\lambda$ the Rashba spin-orbit coupling constant. The Pauli matrices $\bsig$ and $\btau$ act in the spin and particle-hole subspaces, respectively. In momentum space, in the basis $\Psi_{\bs{k}}=(c_{\bs{k},\uparrow},c_{\bs{k},\downarrow},c_{-\bs{k},\downarrow}^\dagger,-c_{-\bs{k},\uparrow}^\dagger)^\mathrm{T}$, the Hamiltonian of this system is given by
\begin{eqnarray}
	\mathcal{H}^{\rm 2D}_{\bs{k}}&=&-\left[\mu+2t\left(\cos k_x+\cos k_y\right)\right]\btau^z\\\nonumber&&
	-2\lambda\left[\sin k_x \bsig^y-\sin k_y \bsig^x\right]\btau^z
	-\Delta\btau^x\,.
\end{eqnarray}
The Shiba chain is modeled as a set of onsite potentials corresponding to local Zeeman fields of the form ${\bm V}=V\bsig^z$, 
\begin{equation}
	\mathcal{H}^{\mathrm{imp}} = \sum\limits_{\bs{r} \in C}\Psi^\dagger_{\bs{r}}{\bm V}\Psi_{\bs{r}}\,,
	\label{eq:HChain}
\end{equation}
with $C$ describing the sites of the one-dimensional infinite chain. Therefore, the full system in real space is described by $\mathcal{H}^{\rm 2D}+\mathcal{H}^{\rm imp}$.


\section{Effective Green's function for the chain} 
The effective one-dimensional Green's function $\mathcal{G}$ for the chain can be computed using the technique presented in Ref.~\cite{Pinon2020}. Thus, for a Hamiltonian $\mathcal{H}^{\rm 2D}_{\bm k}$ we can define the bulk unperturbed Green's function $\mathcal{G}_0(\omega,{\bs{k}})=(\im\omega+\mathcal{H}^{\rm 2D}_{\bm k})^{-1}$. In the presence an impurity potential ${\bf V}\delta(x)$ we can compute an effective Green's function with the help of the $T$-matrix \cite{Byers1993,Salkola1996,Ziegler1996,Mahan2000,Balatsky2006,Bena2016}:
\begin{equation}
{\bf T}=\left[\mathbb{I}-{\bf V}\mathcal{G}_1(\omega,k_y)\right]^{-1}{\bf V}\,,
\end{equation}
where
\begin{equation}
\mathcal{G}_1(\omega,k_y)\equiv\mathcal{G}_1(\omega,x=0, k_y) =\int_{-\pi}^{\pi} \frac{dk_x}{2\pi}\mathcal{G}_0(\omega,{\bs{k}})\,.
\label{eq5}
\end{equation}
In what follows we are interested in evaluating the effective Green's function at $x=0$, i.e. along the impurity chain. In the T-matrix formalism this can be written as:
\begin{equation}
	\mathcal{G}(\omega,k_y)=\mathcal{G}_1(\omega,k_y)+\mathcal{G}_1(\omega,k_y){\bf T}\mathcal{G}_1(\omega,k_y)\,.
	\label{eq:EffGF}
\end{equation}
where $\mathcal{G}_1(\omega,k_y)$ at $\omega = 0$ can be obtained either by performing the integral in Eq.~(\ref{eq5}) numerically, or for the system considered here can also be computed analytically (see Eq.~\eqref{eq:G1ky}) in the appendices). Hereinafter we set $k_y\to k$ where possible. This Green's function defines an effective Hamiltonian for the chain: $\mathcal{H}^{-1}_k\equiv\mathcal{G}(0,k)$. In what follows we apply the three different methods mentioned in the introduction, and calculate the topological phase diagram for the magnetic Shiba chains.


\begin{figure}[t!]
  \includegraphics[height=0.425\columnwidth]{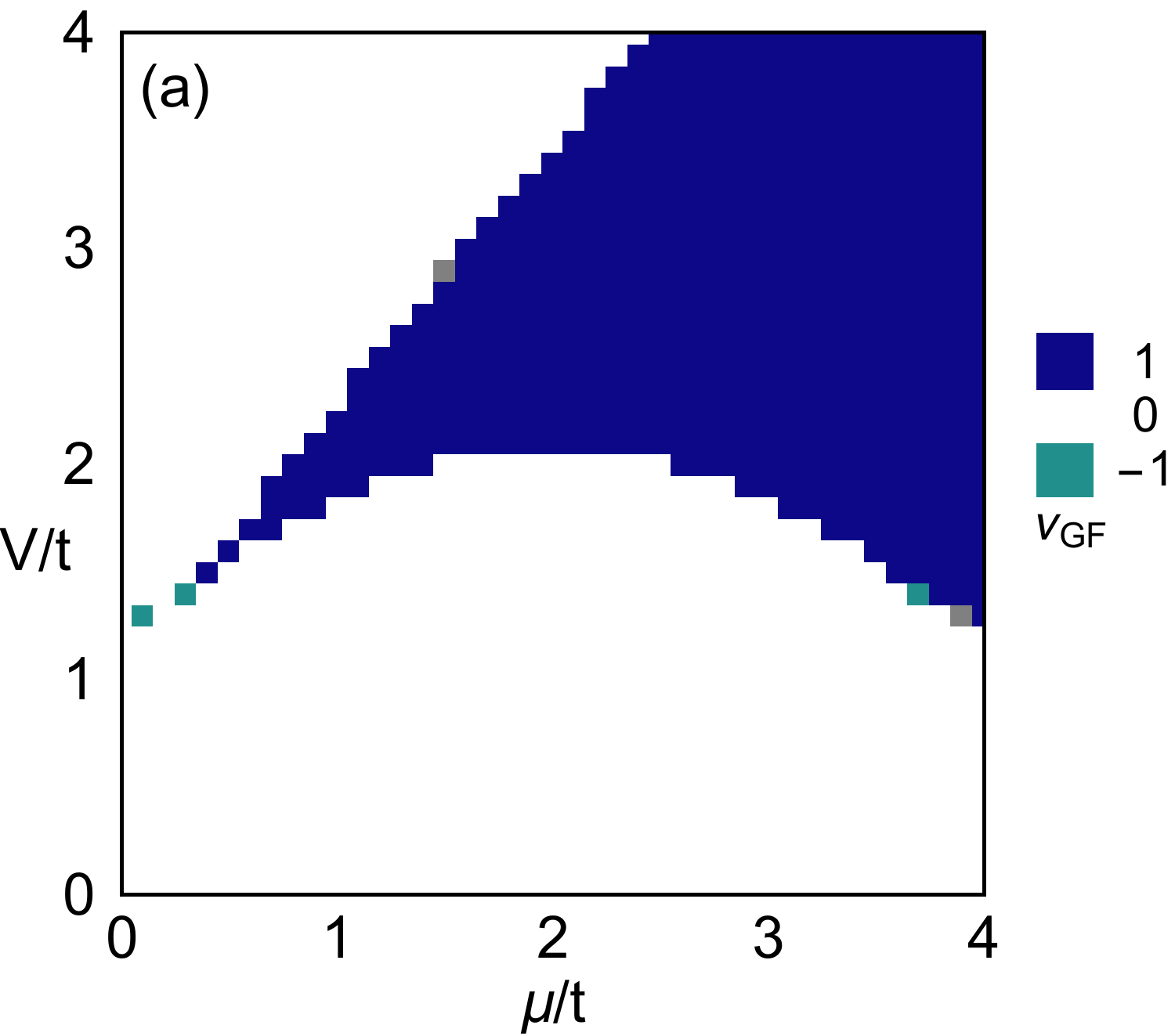}
  \includegraphics[height=0.553\columnwidth]{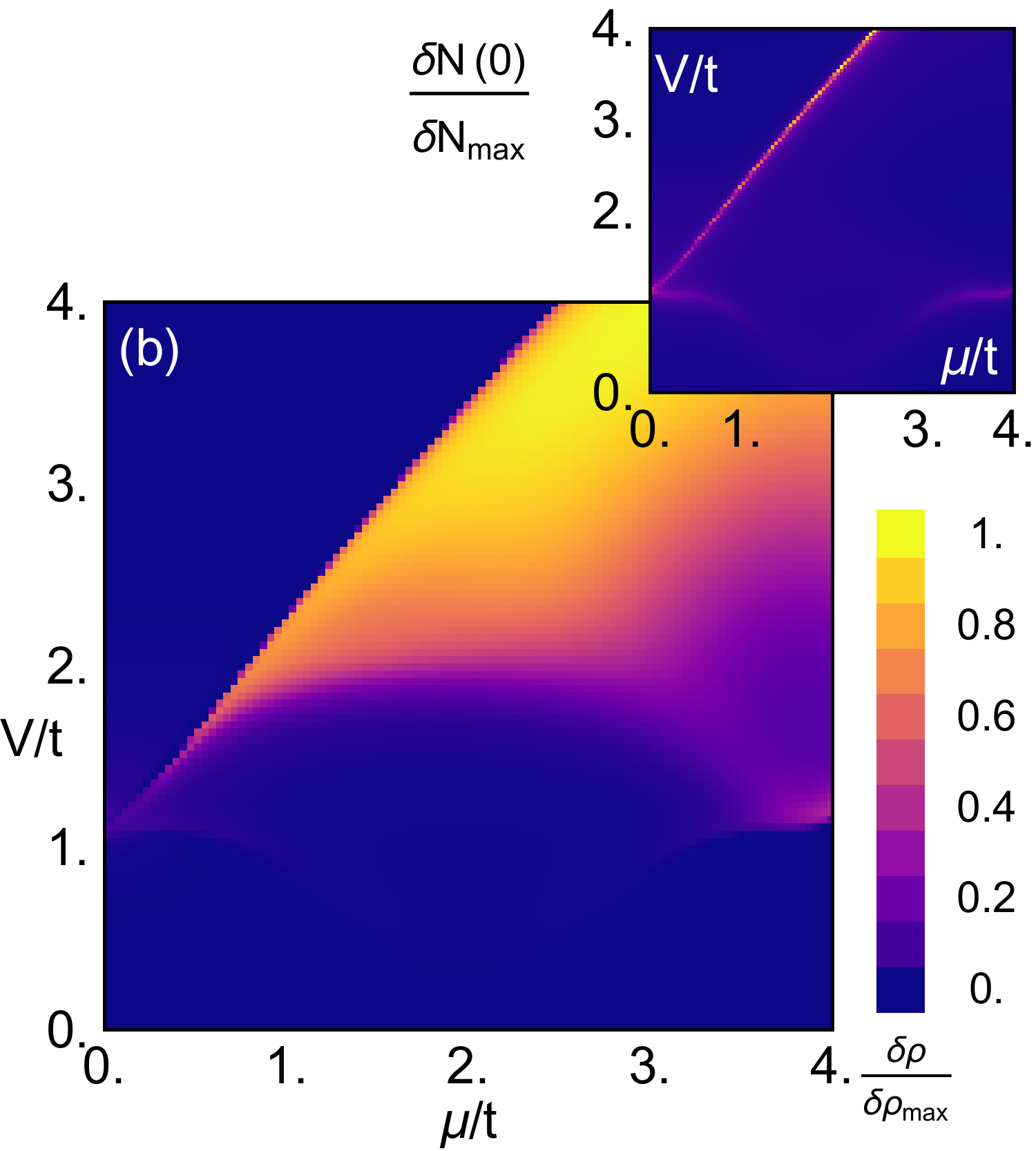}\\
  \includegraphics[height=0.425\columnwidth]{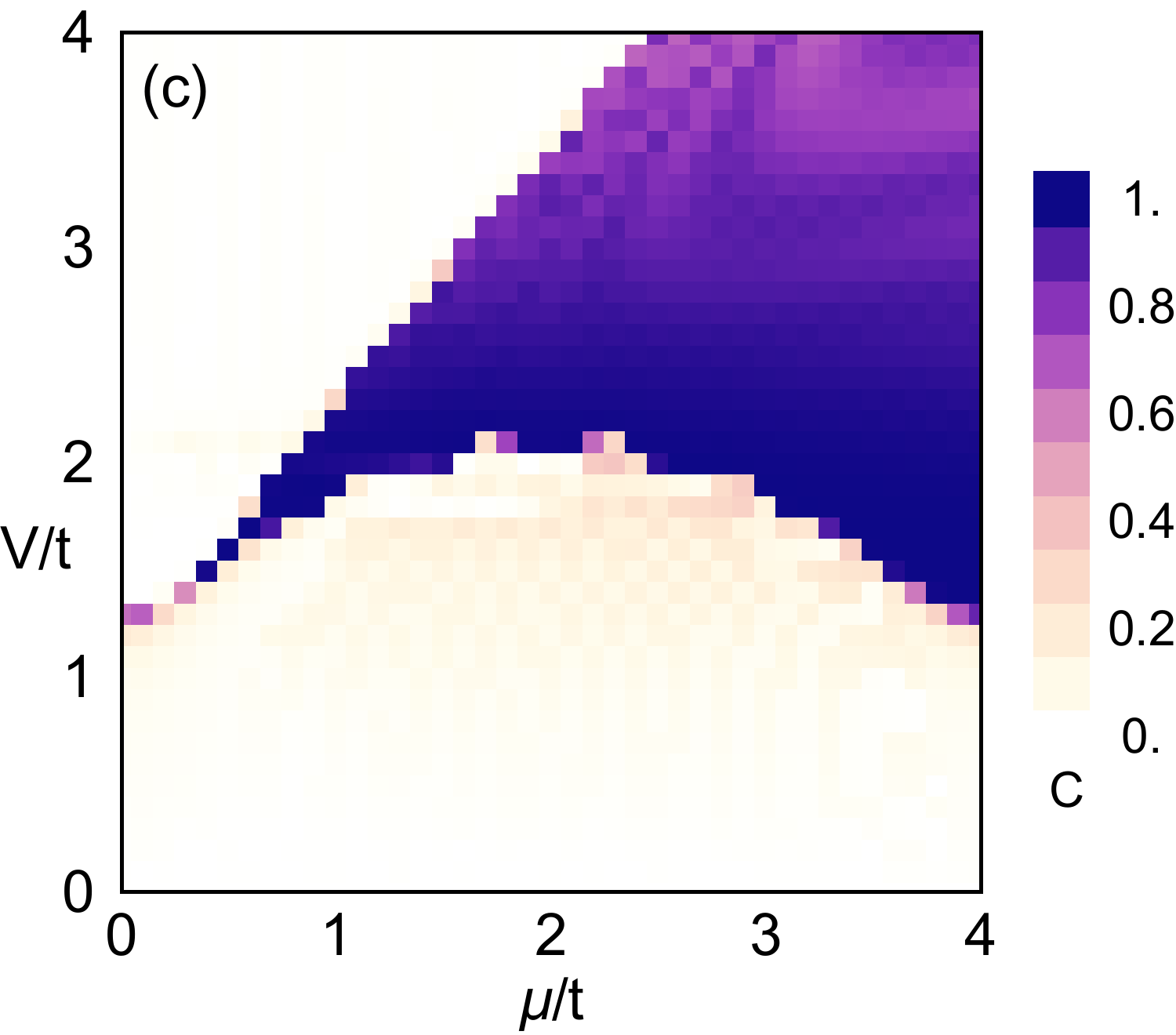}
  \includegraphics[height=0.425\columnwidth]{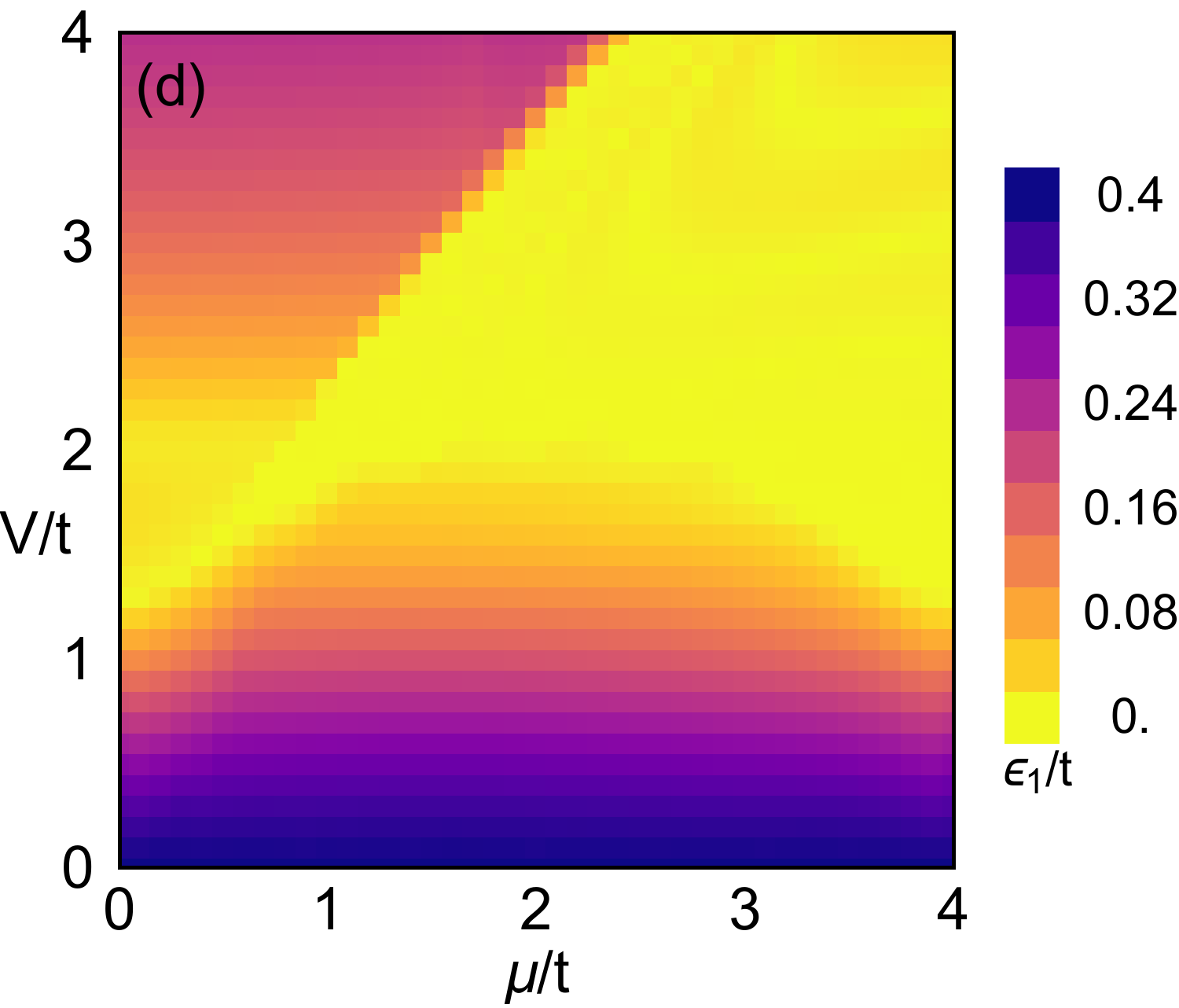}
  \caption{(a) The chiral invariant, $\nu$, plotted as a function of the chemical potential and the magnetic impurity strength, as obtained from the effective wire GF. (b) The zero-energy DOS, $\rho$, in the presence of a probe point-impurity; shown in the inset is the total density of states in the absence of the probe. (c,d) The Majorana polarization $C$ and the energy corresponding to the lowest-energy state calculated using a tight-binding model for a chain of 60 sites introduced into a two-dimensional system of size $21\times80$. We take $\lambda = 0.2t$ and $\Delta= 0.4t$} 
  \label{gapinv1}
\end{figure}


\section{I.~The chiral invariant from the Green's function} In order to calculate the appropriate topological invariant from the effective Green's function we first investigate the symmetries of the problem. The model we consider exhibits several symmetries important for the topology~\cite{Chiu2016,Sato2017}:
	a particle--hole (PH) symmetry described by an anti-unitary operator $\mathcal{P}$, such that $\mathcal{P} \mathcal{H}_{k}^* \mathcal{P}^{-1} = - \mathcal{H}_{-k}$ and $\mathcal{P}^{2} = 1$;
	a ``time-reversal'' (TR) symmetry described by the anti-unitary operator $\mathcal{T}$, where $\mathcal{T} \mathcal{H}_{k}^* \mathcal{T}^{-1} = \mathcal{H}_{-k}$ with $\mathcal{T}^{2} = 1$; and
	finally we have the combination of these symmetries, the sublattice or ``chiral'' symmetry, described by the unitary operator $\mathcal{S} =\mathcal{P}\mathcal{T}$, with $\mathcal{S}^{-1} \mathcal{H}_{k} \mathcal{S} = - \mathcal{H}_{k}$.
When all of these symmetries are present the Hamiltonian is in the BDI symmetry class~\cite{Ryu2010}, and hence has a $\mathbb{Z}$ topological invariant. Our model has a chiral symmetry operator for the effective chain Hamiltonian given by $\mathcal{S}=\btau^y\bsig^x$. Note, however that this is not a symmetry of the full problem, $\mathcal{H}^{\rm 2D}+\mathcal{H}^{\rm imp}$, as the term $\lambda\sin k_x \bsig^y\btau^z$ breaks the TR symmetry. We note here that the two-dimensional lattice described by $\mathcal{H}^{\rm 2D}$ is always topologically trivial. Therefore, we can use the topological invariant~\cite{Gurarie2011},
\begin{equation}
\nu=\frac{1}{4\pi \im}\int_{-\pi}^{\pi}dk\,\Tr\mathcal{S}\mathcal{H}_k\partial_k\mathcal{H}^{-1}_k\,,
\end{equation}
suitable for a one-dimensional system with chiral symmetry.

The result for the chiral invariant as a function of the chemical potential and the magnetic impurity strength for an exemplary set of parameters  ($\Delta=0.4t$ and $\lambda=0.2t$) is given in Fig.~\ref{gapinv1}(a). One can \blue{see} that a large topologically non-trivial phase exists, in particular for large values of $V$.

\section{II.~Testing the formation of Majorana bound states via a probe impurity}
Our second approach is also based on the effective Green's function obtained in Eq.~\eqref{eq:EffGF}. To test the formation of Majorana states we add an infinite-amplitude scalar probe impurity which breaks the wire into two semi-infinite parts, see Fig.~\ref{fig:systemsketch}(b) \cite{Kaladzhyan2019,Zazunov2016}. The perturbed Green's function of the wire in the presence of the probe impurity can be obtained via the $T$-matrix formalism using the effective Green's function of the infinite wire in Eq.~(\ref{eq:EffGF}) as an unperturbed Green's function describing the infinite chain. Assuming that the probe impurity is localized at $y=0$ and is described by $U \mathrm{diag}\, \{1, 1, 1, 1 \} \delta(y)$, we can find the $T$-matrix and the correction to the Green's function as follows:
\begin{align}
\tilde{T} = \left[\mathbb{I} - U \int_{-\pi}^{\pi} \frac{dk}{2\pi} \mathcal{G}(0,k) \right]^{-1} U
\end{align}
 The local density of states is given by
\begin{align}
\delta \tilde{\rho}(0, y) = -\frac{1}{\pi} \Imag\, \Tr\, \mathcal{G}(0,y) \tilde{T} \mathcal{G}(0,-y),
\end{align}
where $\mathcal{G}(0,y) = \int_{-\pi}^\pi \frac{dk}{2\pi} \mathcal{G}(0,k) e^{\im k y}$. To verify the formation of Majorana states, in Fig.~\ref{gapinv1}(b) we plot the zero-energy local density of states in the presence of the infinite impurity, as a function of the chemical potential and the magnetic impurity strength in the chain. To check that these states are indeed forming inside a bulk gap and are not just regular bulk states, in the inset we plot the zero-energy total DOS in the bulk of the wire:
\begin{align}
\delta N(0) = \int dy\, \delta \tilde{\rho}(0, y)\,.
\end{align}
We confirm that this is indeed negligible except along the bulk-gap closing lines. Thus the states that we observe at zero-energy do not correspond to any bulk states, but come from adding the impurity and are thus Majorana bound states. This is in very good agreement with the chiral invariant in Fig.~\ref{gapinv1}(a).

We stress here that the phase diagram we have calculated can not be directly compared to any in the literature which typically consider the dilute limit of Shiba impurities, for example Refs.~\cite{Pientka2013,Brydon2015,Heimes2015,Weststrom2015,Schecter2016,Mashkoori2020}.


Our results are checked against standard tight-binding calculations, thus in Figs.~\ref{gapinv1}(c,d) we consider a finite-size system of dimensions $21\times80$, with a magnetic impurity chain embedded in it, and we solve the tight-binding model in real space numerically. We focus on the lowest-energy state, and we calculate the corresponding Majorana polarization $C$\cite{Sedlmayr2015b,Sedlmayr2016,Sedlmayr2017,Kaladzhyan2017a,Kaladzhyan2018,Glodzik2020}, i.e., the normalized expectation value of the particle-hole operator integrated over half the wire. For a localized Majorana bound state, i.e.~an eigenstate of $\mathcal{P}$, we should have $C=1$, and this is indeed what we observe in Fig.~\ref{gapinv1}(c). We note that there is a perfect agreement between our two new methods introduced above, and the numerical tight-binding analysis: the topological regions in the phase diagram as predicted by our methods are exactly the same as those that exhibit a perfect Majorana polarization ($C=1$) and a zero-energy state in the numerical tight-binding calculation.


\section{III.~Invariant from the quasi-one-dimensional strip}
We now consider that the system described by $\mathcal{H}^{\rm 2D}+\mathcal{H}^{\rm imp}$ has a finite number of sites along the $x$-direction, $N_x$. The resulting system is a quasi-1D strip described by the Hamiltonian $\mathcal{H}^{\rm Q1D}$, with the finite-size segment in the $x$-direction constituting the unit cell. This configuration breaks TR symmetry and so the system belongs to the class D with a $\mathbb{Z}_2$ invariant. This invariant can be calculated by considering the parity of the bands at high symmetry points, which encodes the band inversion~\cite{Sato2009b}. At the time reversal invariant momenta, following a rotation the Hamiltonian can be written as $\mathcal{H}^{\rm Q1D}_{0,\pi}\to{\rm diag}\,(\bar{\mathcal{H}}_{0,\pi},-\bar{\mathcal{H}}_{0,\pi})$, and calculating the parity of the negative energy bands becomes equivalent to calculating $\delta=(-1)^{\nu_{\rm Q1D}}=\sgn\left[\det\bar{\mathcal{H}}_0\det\bar{\mathcal{H}}_\pi\right]$. Following Ref.~\cite{Sedlmayr2016} a closed form expression can be found for $\delta$. As $\nu_{\rm Q1D}$ is a $\mathbb{Z}_2$ invariant we can unambiguously write that
\begin{equation}\label{invariant}
\nu_{\rm Q1D}=\frac{1-\sgn\left[\det\bar{\mathcal{H}}(0)\det\bar{\mathcal{H}}(\pi)\right]}{2}\,.
\end{equation}
More details can be found in appendix ref{appinv}.

\begin{figure}[h]
  \includegraphics[height=0.425\columnwidth]{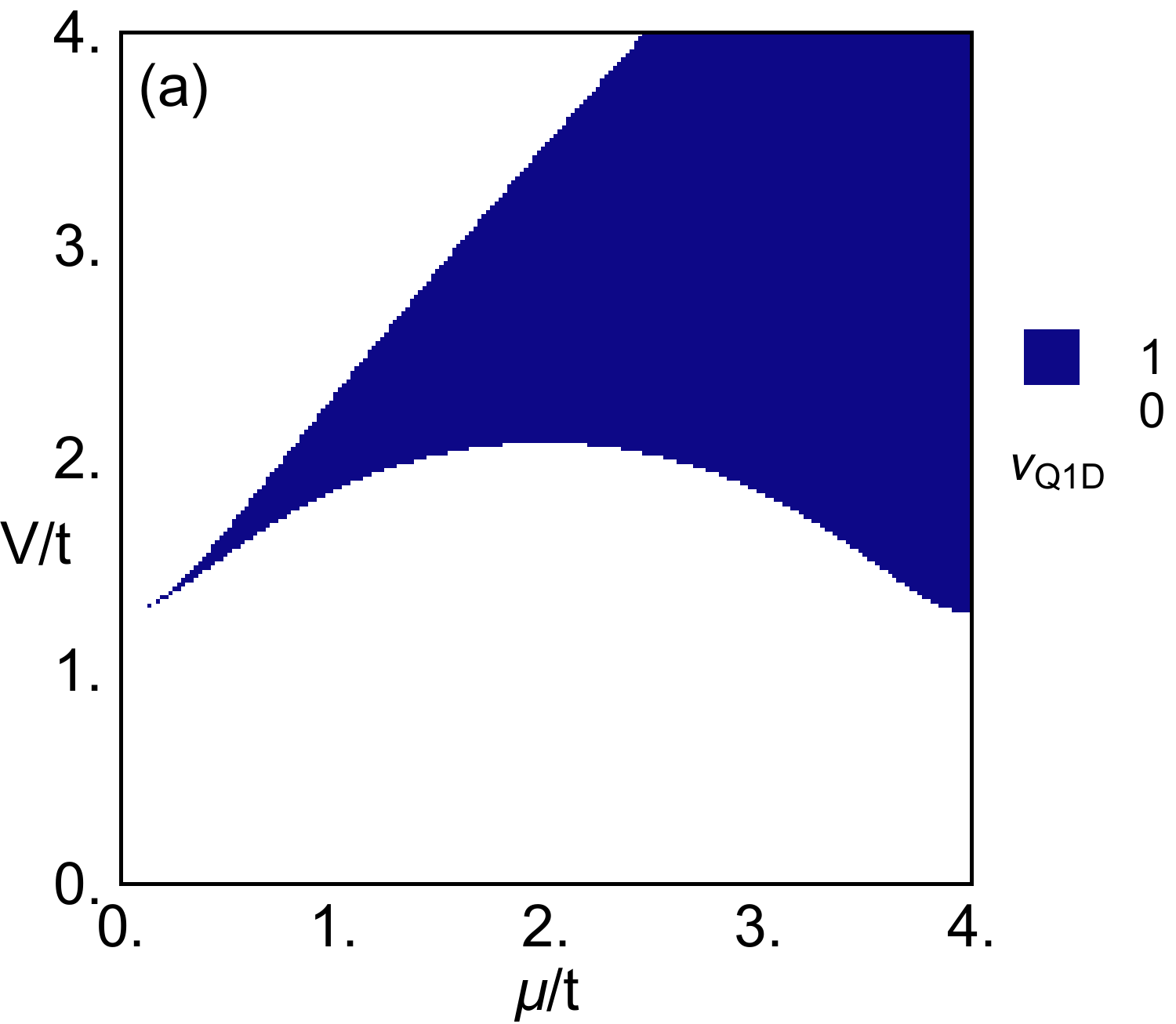}
   \includegraphics[height=0.425\columnwidth]{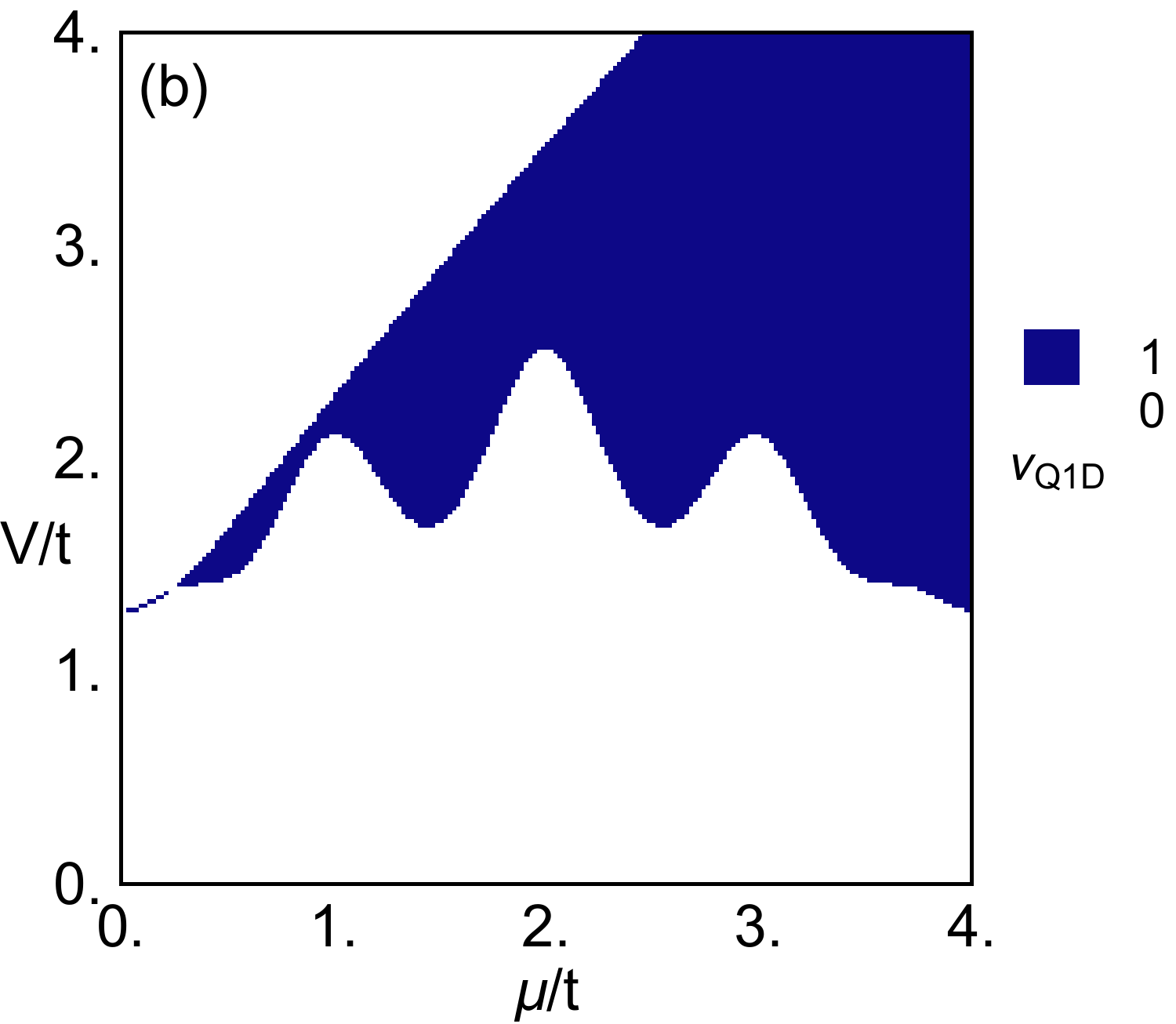}
  \caption{The quasi-one-dimensional invariant, $\nu_{\rm Q1D}$, for (a) $N_x=51$ and (b) $N_x=11$.}\label{fig3}
\end{figure}

The quasi-one-dimensional invariant works very well, and its predictions agree perfectly with the previous calculations for the topological phase diagram presented in Fig.~\ref{gapinv1}, provided $N_x$ is large enough. In Fig.~\ref{fig3}(a) we plot the quasi-one-dimensional invariant as a function of chemical potential and magnetic impurity strength for a a strip of width of $N_x=51$, while in the right panel we consider $N_x=11$. Note that the oscillations present at small widths go away for $N_x=51$.

\section{Comparison with experiments} To facilitate the numerical computations, so far we have used unrealistically large values of the parameters. In what follows we consider a set of parameters expected to correspond more closely to realistic experimental parameters. In the experiments the value of the superconducting gap varies from $1.1$ to $1.4$~meV  \cite{Nadj-Perge2014,Pawlak2015,Ruby2015,Feldman2017a,Ruby2017,Kim2018b}, while the estimate for the spin-orbit coupling constant and magnetic exchange constant in Ref.~\cite{Nadj-Perge2014} are $22$~meV \footnote{The value provided in the aforementioned reference is $0.11$~eV$\cdot$\AA. To convert it into the units of energy, we divide this value by the lattice constant of lead ($a = 4.95$~\AA), and we get $\lambda \approx 22$~meV.} and $2.4$~eV, respectively.

\begin{figure}
	\begin{minipage}[c]{0.6\linewidth}
	\includegraphics[width=\linewidth]{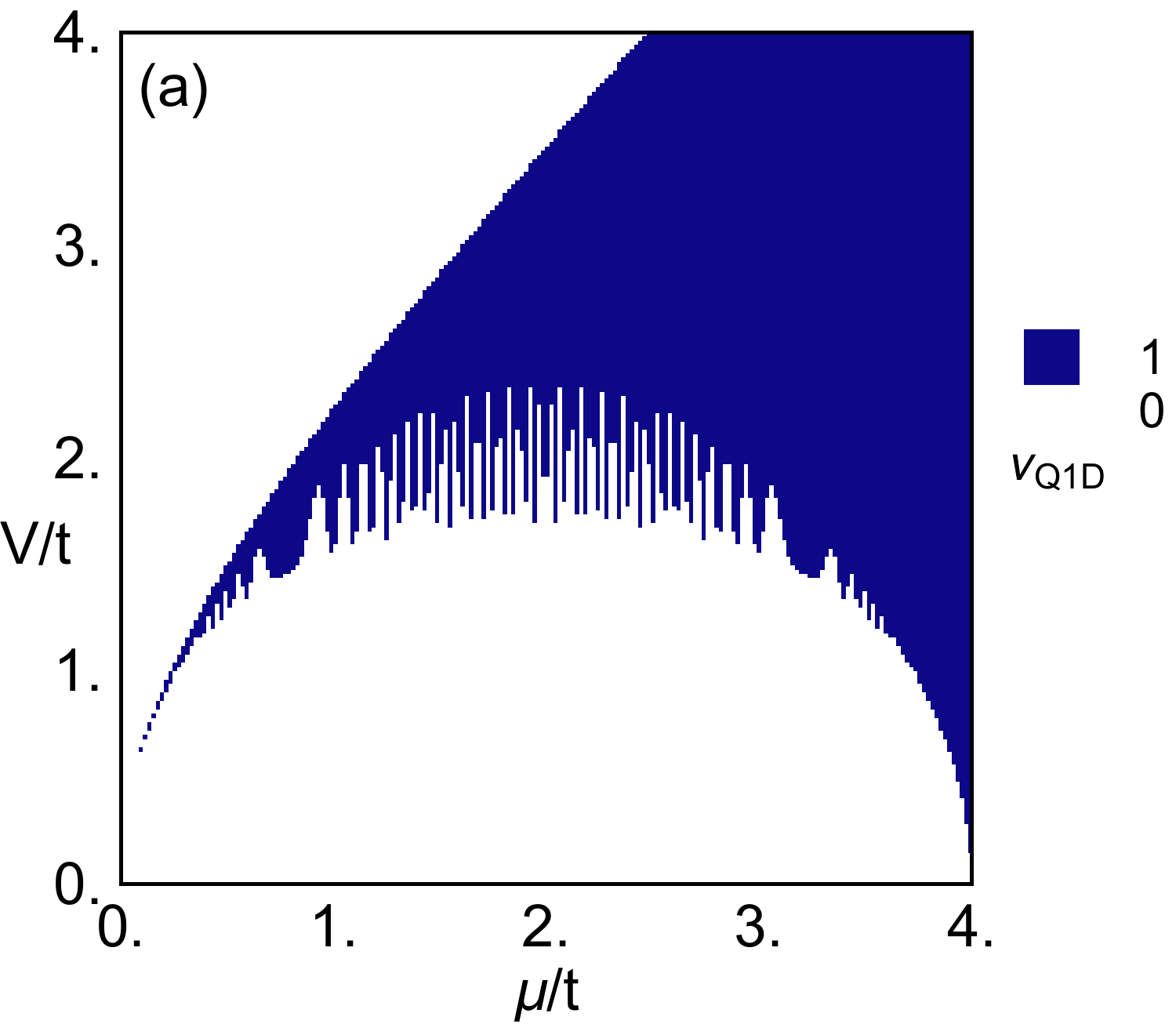}
	\end{minipage}
	\begin{minipage}[c]{0.3\linewidth}
		\includegraphics[height=1\linewidth]{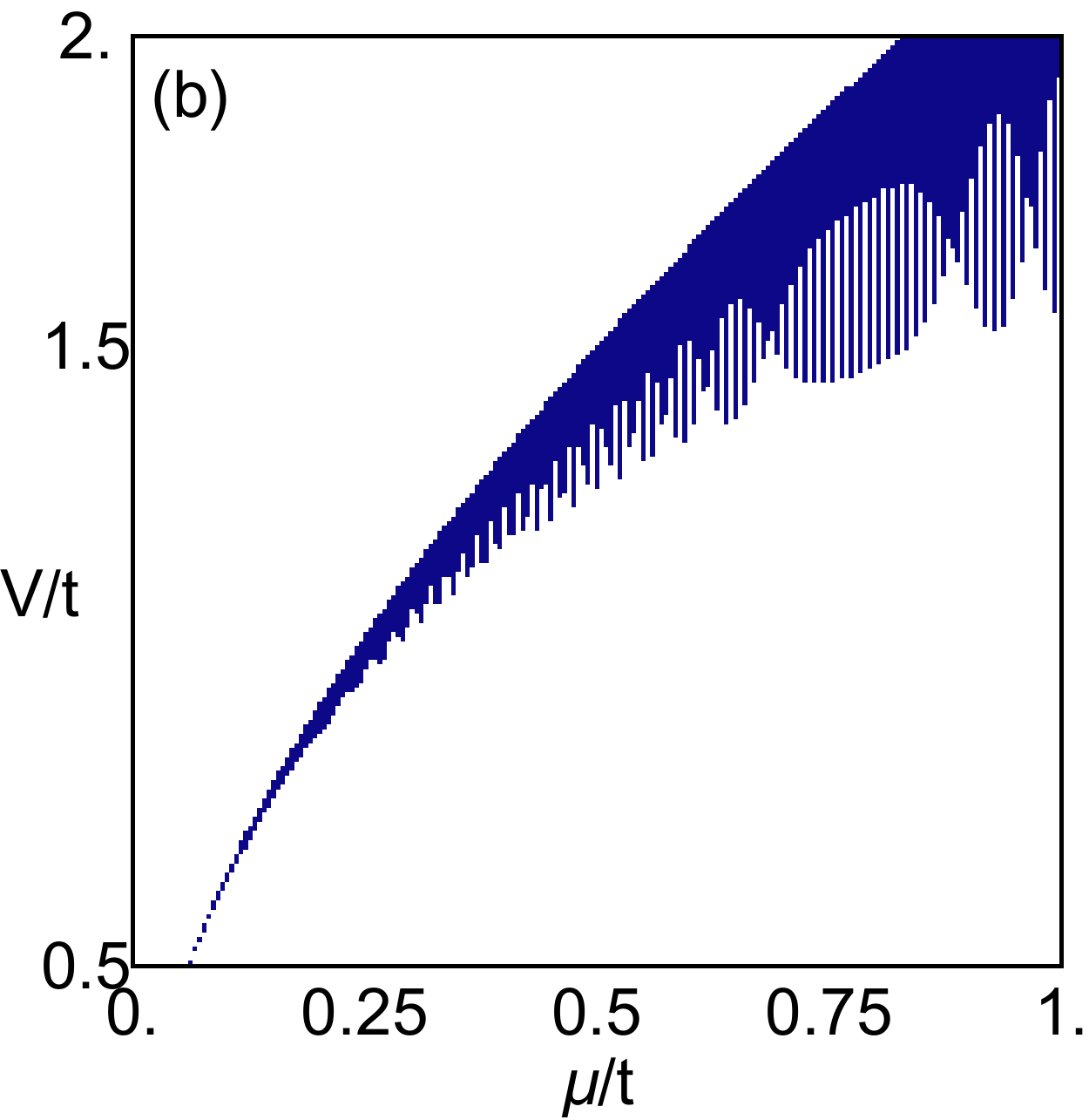}\\
		\includegraphics[height=1\linewidth]{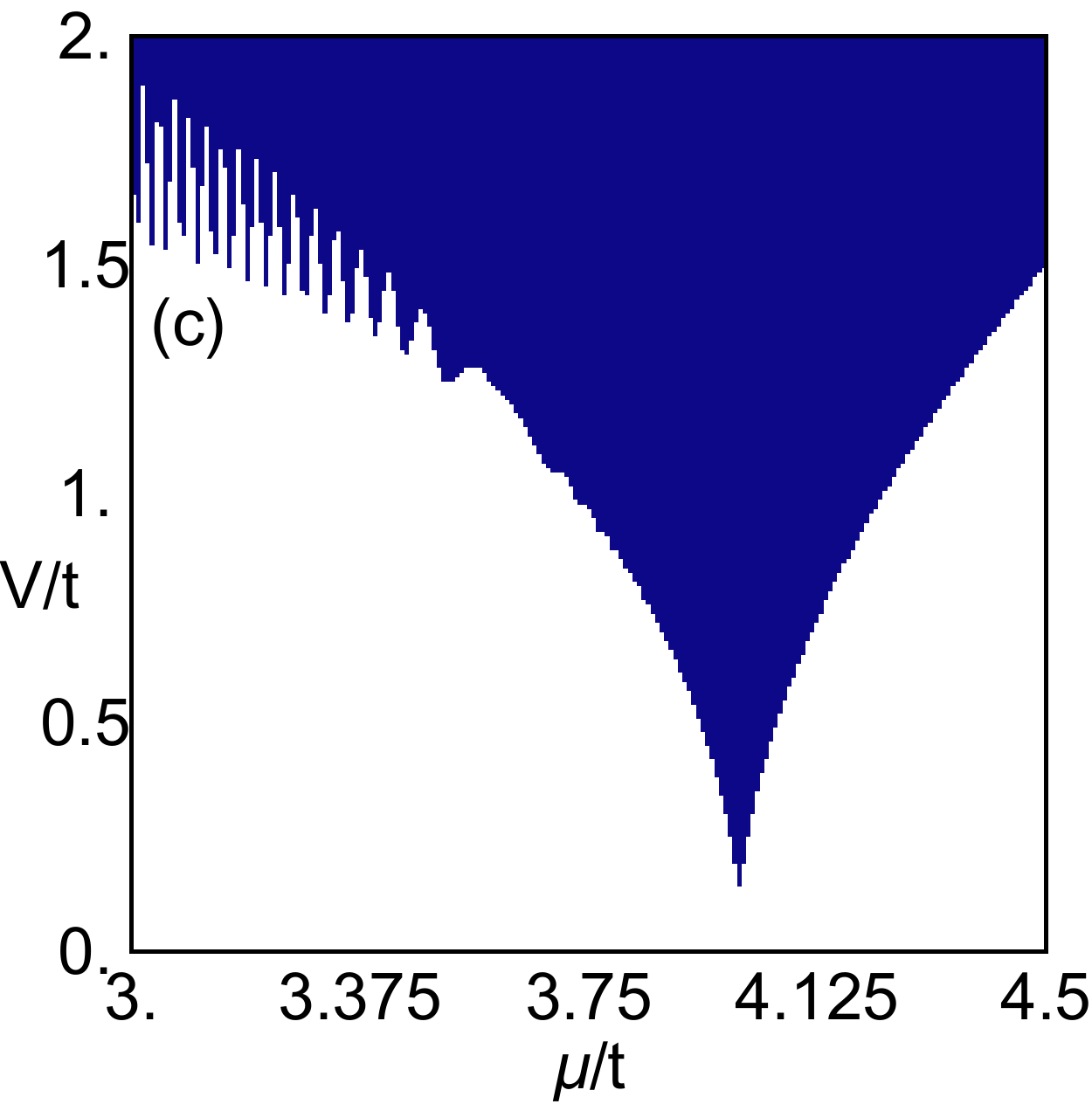}
	\end{minipage}
	\caption{The quasi-one-dimensional invariant for $\lambda = \Delta=0.005t$ for  $N_x=1001$. Small panels (b,c) demonstrate the parts of the phase diagram close to $\mu/t = 0$ and $\mu/t = 4$ in a higher resolution.}
	\label{fig:gapinv2}
\end{figure}

In Fig.~\ref{fig:gapinv2} we plot the quasi-one-dimensional invariant for  $\lambda = \Delta=0.005t \sim 8$~meV, closer to what one may expect in a realistic system. We note that the convergence is slower in this case with the width of the system, but the results are comparable to those for larger $\Delta$ and $\lambda$. The main differences come from the points close to $\mu=0$ and  $\mu=4$ and for these points we performed detailed analyses in the panels (b,c) of Fig.~\ref{fig:gapinv2}. We note in particular that smaller values of the magnetic impurity potential are required close to $\mu=4$ to give rise to a topological phase. It appears that quite large magnetic impurity potentials are necessary in the other regions of the phase diagram, which may not be always easy to achieve experimentally. However, we note that this could be achieved in an alternative configuration, in which impurities are deposited above the surface atoms and connected with the substrate via a tunneling coupling $t'$. We found that this configuration of suspended atoms with an impurity potential $U$ is exactly equivalent with that of embedded impurities with an effective impurity potential $V=t'^2/U$. Thus larger values of the effective impurity potential $V$ can be reached, even for not so large values of the bare $U$, see appendix \ref{adatoms} for more details.

\section{Conclusions}
We proposed three new analytical and semi-analytical techniques to calculate the topological phase diagrams of Shiba chains. The first relies on calculating the the effective Green's function of a {\it line impurity} using the T-matrix formalism and evaluating the corresponding chiral invariant. The second technique is also based on the effective GF of the chain, and consists in calculating the zero-energy DOS in the presence of a probe impurity. The third method is based on calculating the quasi-one-dimensional invariant of a ribbon with an embedded impurity chain in the limit of infinite ribbon width. We applied our techniques to the example of a chain of magnetic impurities on an $s$-wave superconducting substrate with Rashba spin-orbit coupling. We compared our results to those obtained via tight-binding numerical calculations, and we found a perfect agreement. We discussed the experimental relevance of our results. Our techniques are very general and it would be useful to apply them to other problems, especially some that are difficult to solve correctly numerically due to finite size effects \cite{Stanescu2011a,Peng2015}.


\acknowledgments

This work was supported by the National Science Centre (NCN, Poland) under the grant 2019/35/B/ST3/03625 (NS).

\appendix

\section{Quasi-one-dimensional invariant}\label{appinv}

We can write the Hamiltonian for $\mathcal{H}^{\rm 2D}+\mathcal{H}^{Ch}$ as if the $x$ direction is a unit cell, and making a Fourier transform along $y$:
\begin{equation}\label{kham}
\mathcal{H}^{\rm Q1D}_k=\begin{pmatrix}
 {\bm f}_{ k}+\mathcal{V}& \CuL_{ k} & -\Delta & 0\\
\CuL^\dagger_{ k}& {\bm f}_{ k}-\mathcal{V} & 0 & -\Delta\\
 -\Delta  &0 & \mathcal{V}-{\bm f}^\dagger_{- k} & \CuL^T_{- k} \\
0&  -\Delta &  \CuL^*_{- k}& -{\bm f}^\dagger_{- k}-\mathcal{V}
\end{pmatrix}\,.
\end{equation}
We write $\mathcal{V}=V\mathcal{D}_{C_y}$ where $\mathcal{D}_{C_y}$ is a square matrix with a non-zero entry only for the site of the chain $C_y$, and $\Delta$ is a  diagonal matrix. Furthermore we have
\begin{equation}
{\bm f}_k=-[2t\cos[k]+\mu]\mathbb{I}_{N_x}-t\mathbb{I}^+_{N_x}-t\mathbb{I}^-_{N_x}\,,
\end{equation}
and
\begin{equation}
\CuL_{ k}=-2\im\alpha\sin[k]\mathbb{I}_{N_x}-\im\alpha\mathbb{I}^+_{N_x}+\im\alpha\mathbb{I}^-_{N_x}\,,
\end{equation}
where $\mathbb{I}_{N_x}$ is an $N_x\times N_x$ identity matrix and $[\mathbb{I}^\pm_{N_x}]_{ij}=\delta_{i,j\pm1}$ are also $N_x\times N_x$ matrices.

We write $\tilde{\mathcal{H}}( k)=\mathcal{U}^\dagger\mathcal{H}( k)\mathcal{U}$ where
\begin{equation}
\mathcal{U}=\frac{1-{\bm\sigma}^y{\bm\tau}^y}{2}\mathbb{I}_{N_x}+\frac{{\bm\sigma}^z{\bm\tau}^z-{\bm\sigma}^x{\bm\tau}^x}{2}\bar{\mathbb{I}}_{N_x}\,,
\end{equation}
we have introduced an $N_x\times N_x$ matrix whose entries are $\left[\bar{\mathbb{I}}_{N_x}\right]_{nn'}=\delta_{n,N_yx+1-n'}$. Then
\begin{equation}
\tilde{\mathcal{H}}(\hat\Gamma_i)=\begin{pmatrix}
\bar{\mathcal{H}}(\hat\Gamma_i)&0\\
0&-\bar{\mathcal{H}}(\hat\Gamma_i)
\end{pmatrix}\,,
\end{equation}
from which we can calculate the invariant same as in the main text~\cite{Sato2009b,Sedlmayr2016}:
\begin{equation}\label{invariantsm}
\nu_{\rm Q1D}=\frac{1-\sgn\left[\det\bar{\mathcal{H}}(0)\det\bar{\mathcal{H}}(\pi)\right]}{2}\,.
\end{equation}
For example for $N_x=3$ we find, with $t=1$,
\begin{align}
	\det\bar{\mathcal{H}}_{0,\pi}=&-\left(\Delta^2+(\mu\pm2)^2\right)
	\nonumber\\&
	\times\left[\Delta^4-V^2\left(\Delta^2+(\mu\pm2)^2\right)
	\nonumber\right.\\&\left.
	+\left(2-2\lambda^2+\mu(\mu\pm4)\right)^2
	\nonumber\right.\\&\left.
	+2\Delta^2\left(6+2\lambda^2+\mu(\mu\pm4)\right)
	\right]\,.
\end{align}
For larger $N_x$ the expressions become rapidly rather complicated, but can be calculated numerically relatively quickly.

It is worth noting that the effective 1D chain calculated from the Green's function and the quasi-1D system considered here belong to different symmetry classes with different invariants, $\mathbb{Z}$ and $\mathbb{Z}_2$ respectively. In the case in which the chiral invariant $|\nu|\geq2$ one can ask what happens to the additional Majorana bound states in the quasi-1D picture. They would become gapped on an energy scale related to the inverse transverse width of the strip. For wide enough strips one would neglect such an effect. We note that this situation does not occur for the model we are considering.
\vspace{.1in}

\section{Impurities as a suspended chain}\label{adatoms}

\begin{figure}
	\centering
	\includegraphics[width=0.475\columnwidth]{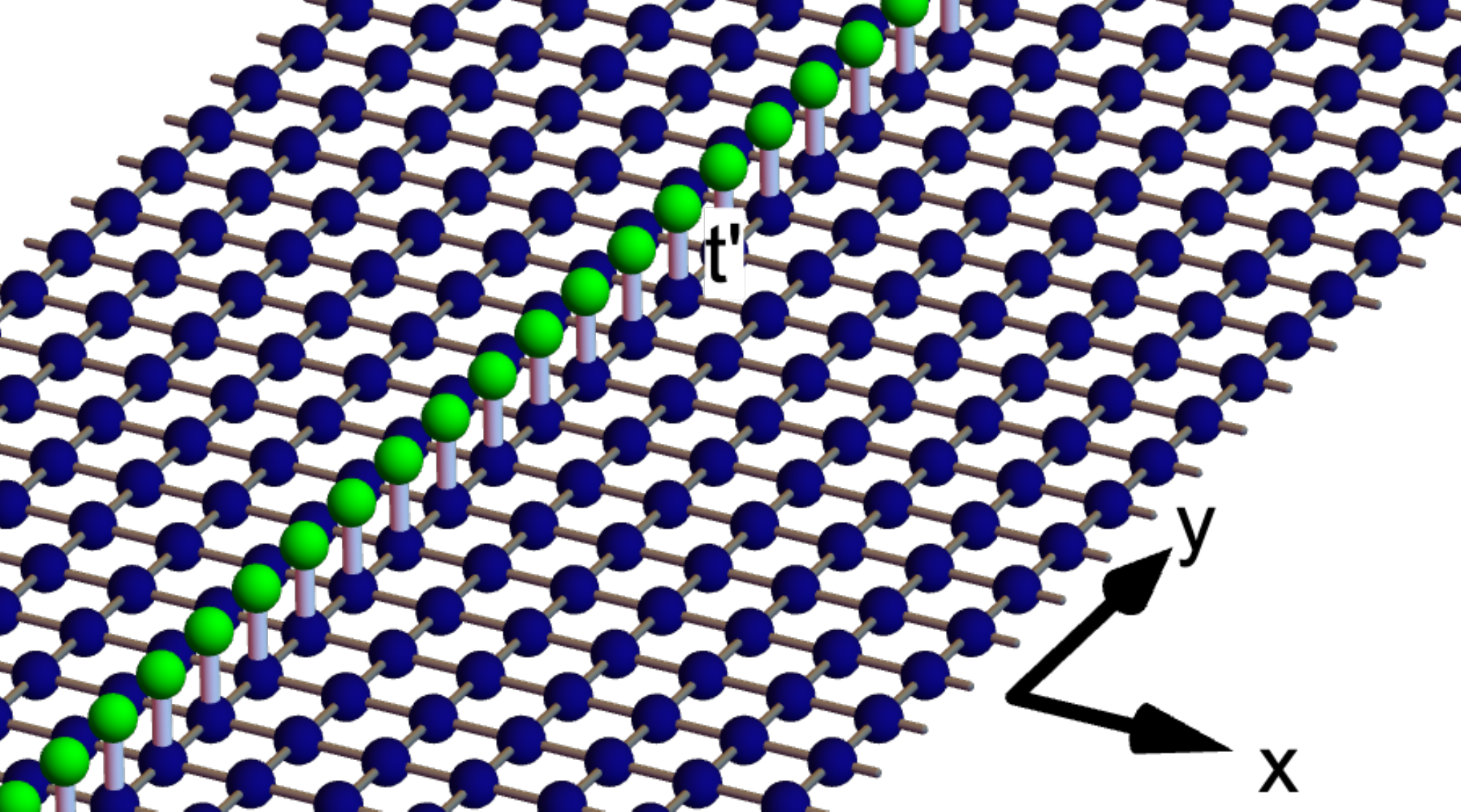}
	\caption{A schematic of the magnetic impurity sites (green atoms) above the superconducting surface (dark blue atoms).}
	\label{fig:systemsketchsm}
\end{figure}

If the magnetic impurity sites for the chain are above the superconductor's surface, see Fig.~\ref{fig:systemsketchsm}, then the system is described by the Hamiltonian $\mathcal{H}^{\rm 2D}+\mathcal{H}^{\rm Ad}$ where 
\begin{equation}
	\mathcal{H}^{\rm Ad}=t'\sum_{\mathbf{r}\in C}\Psi^\dagger_{\mathbf{r}}{\bm\tau}^z\Phi_{\mathbf{r}}+\textrm{H.c.}
	+U\sum_{\mathbf{r}\in C}\Phi^\dagger_{\mathbf{r}}{\bm\sigma}^z\Phi_{\mathbf{r}}\,.
\end{equation}
$\Phi_{\bs{r}}=(a_{\bs{r},\uparrow},a_{\bs{r},\downarrow},a_{\bs{r},\downarrow}^\dagger,-a_{\bs{r},\uparrow}^\dagger)^\mathrm{T}$, where $a_{\bs{r},\sigma}^{(\dagger)}$ annihilates (creates) an electron with spin $\sigma = \uparrow, \downarrow$ at site $\bs{r}\in C$. $\mathcal{H}^{\rm 2D}$ is given by \eqref{eq:H2DSC}, same as in the main text.

\begin{figure}[h]
  \includegraphics[height=0.4\columnwidth]{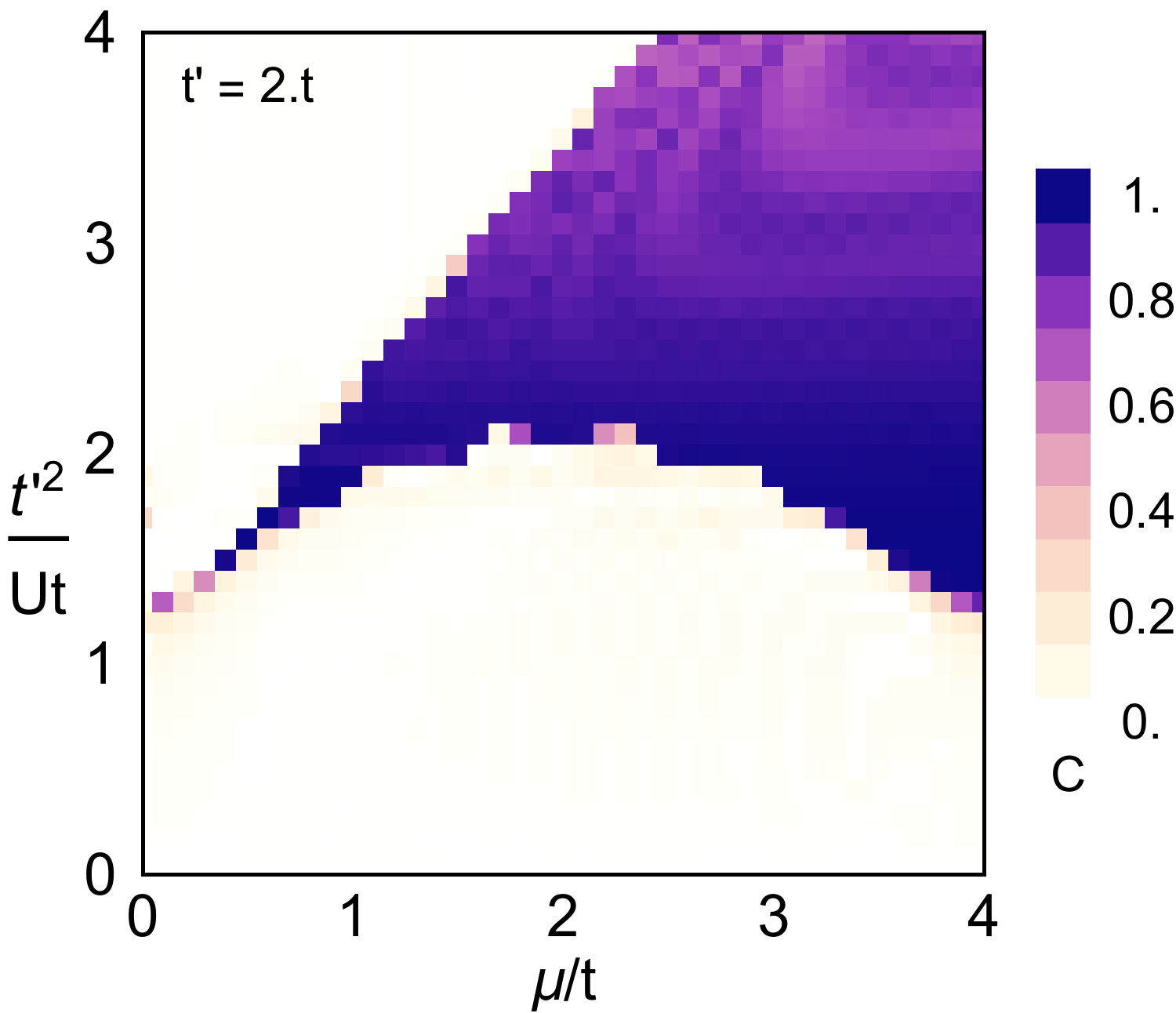}\hspace{0.1cm}
  \includegraphics[height=0.4\columnwidth]{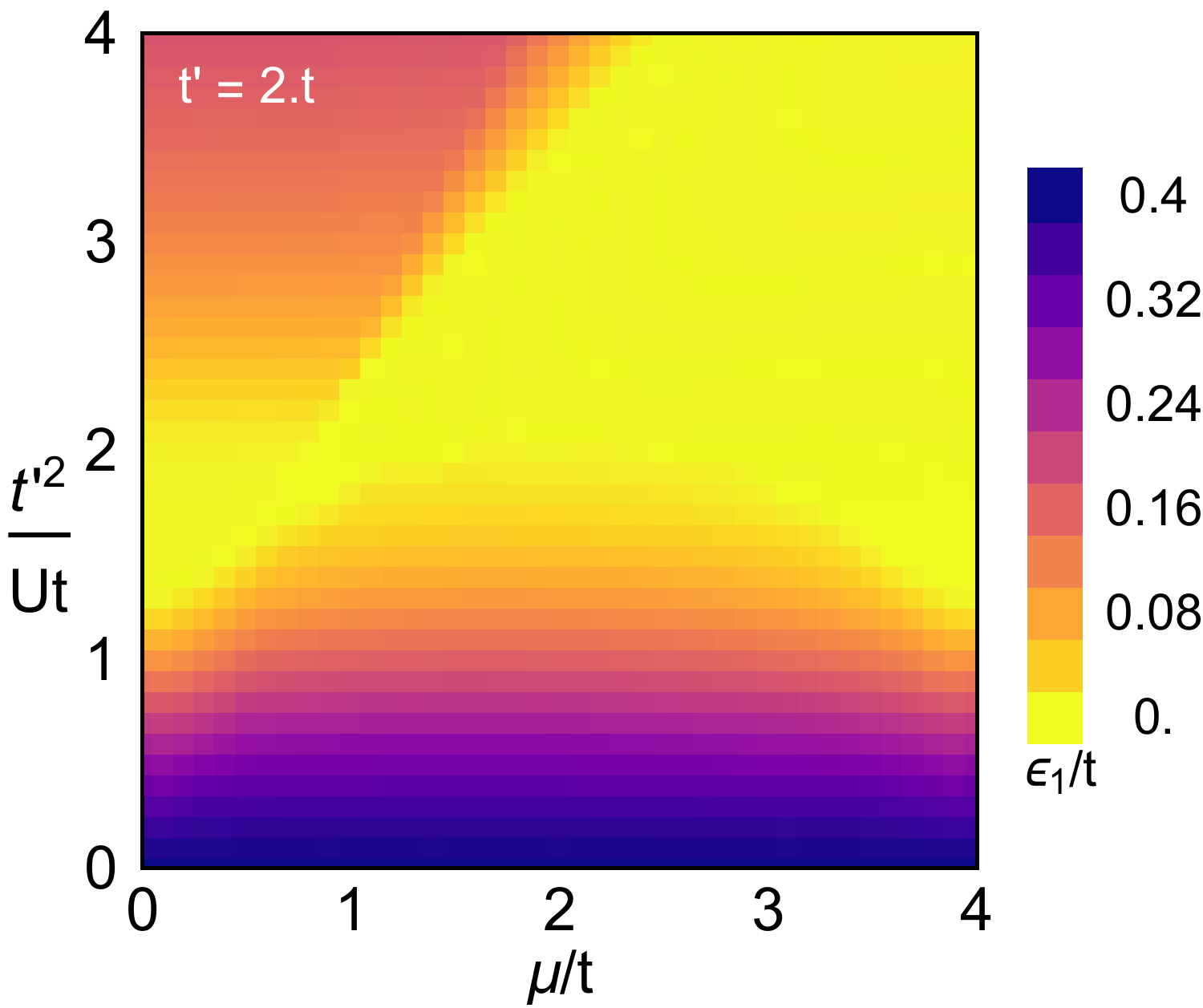}\hspace{0.1cm}\\
  \includegraphics[height=0.4\columnwidth]{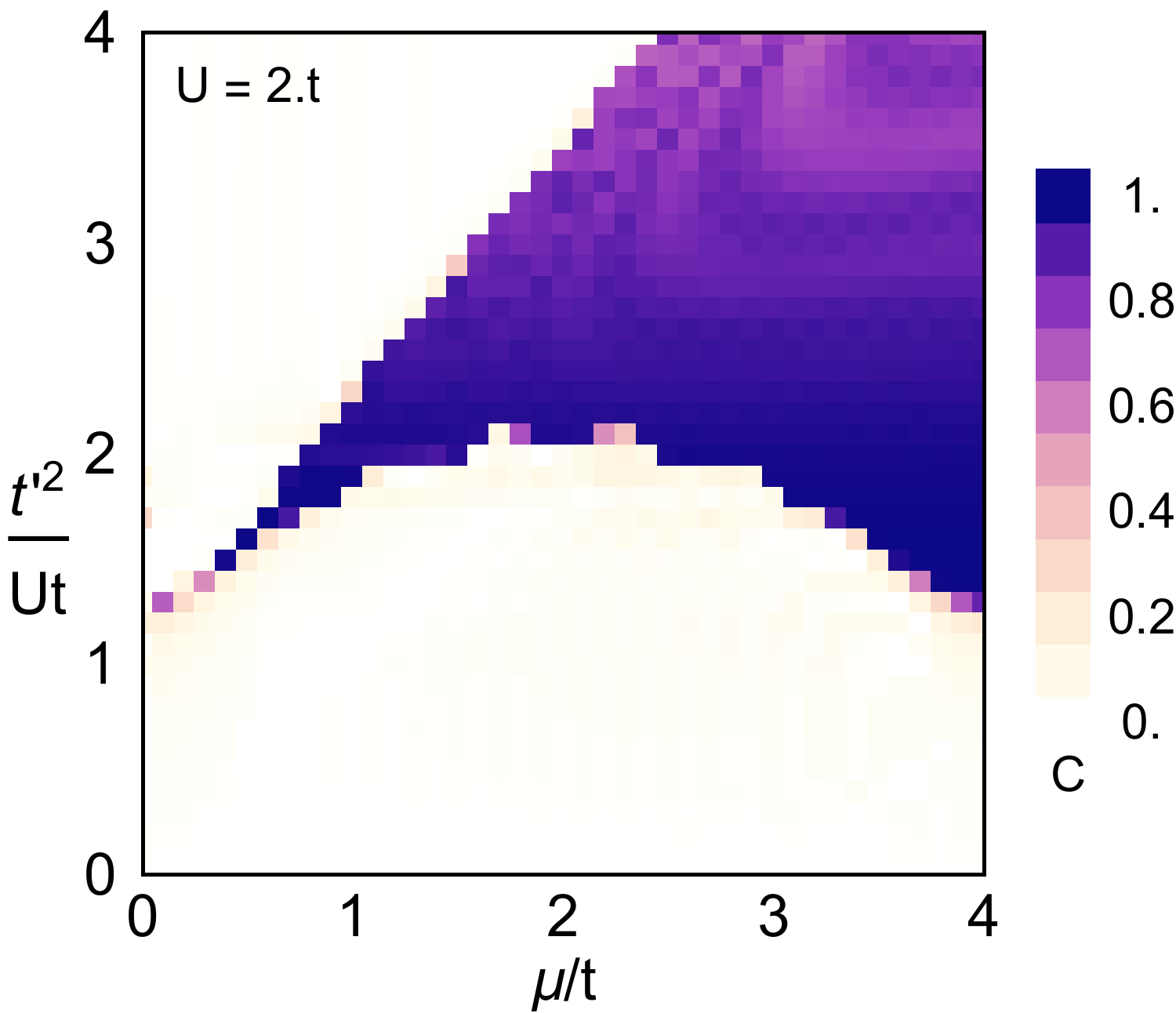}\hspace{0.1cm}
  \includegraphics[height=0.4\columnwidth]{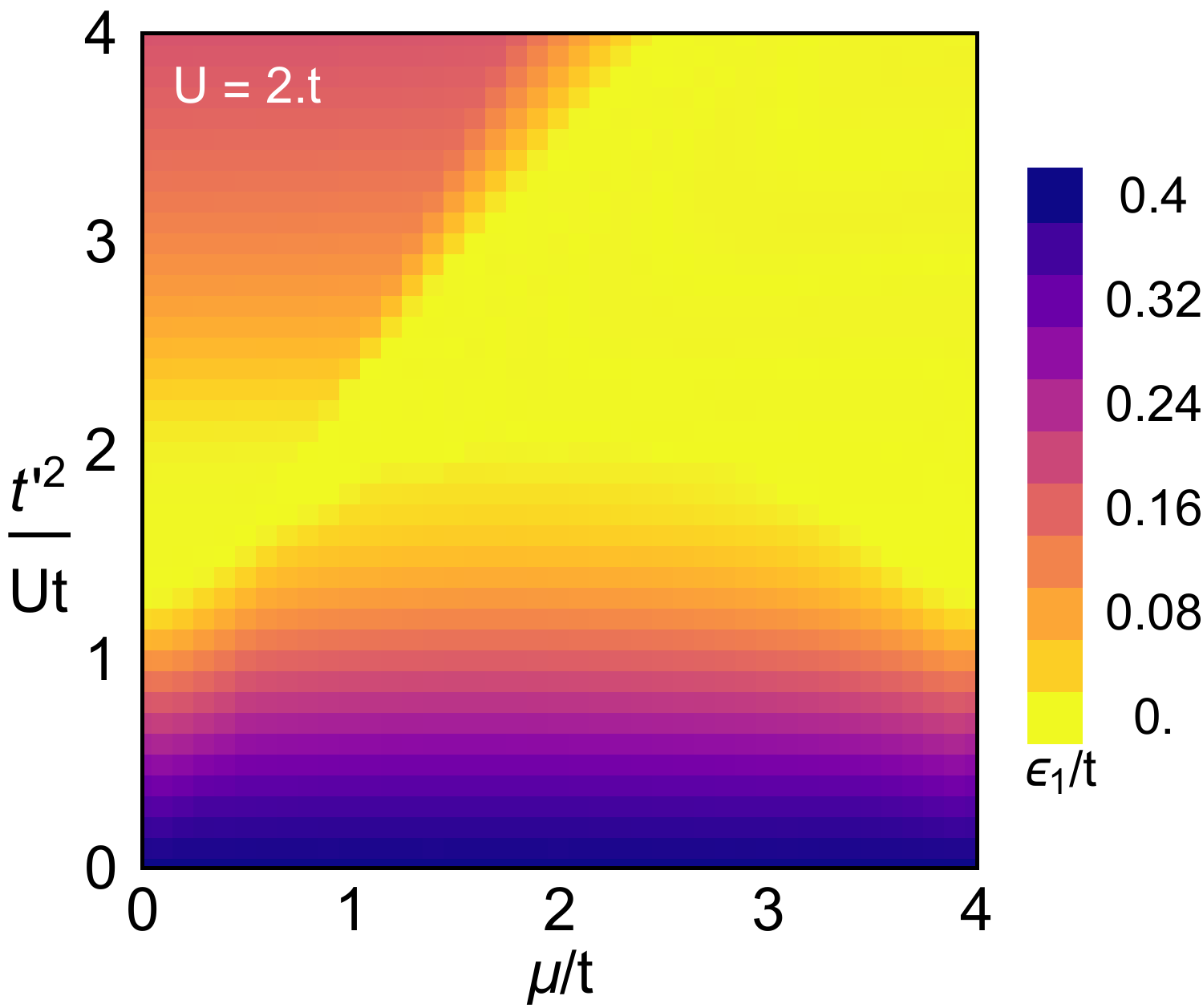}
  \caption{A comparison of the lowest energy state energy $\epsilon_1$ and its Majorana polarization $C$ for the impurities defined as sites {\it above} the two-dimensional lattice. Here we take $\lambda = 0.2t$, and $\Delta= 0.4t$. Either $t'$ or $U$ are kept constant and the other term is varied, as labelled in the figures.}    
  \label{figs}
\end{figure}

 The effect of the extra impurity sites is to generate an effective Zeeman term on the sites directly underneath given by $V=t'^2/U$. One can therefore approximately map this system to $\mathcal{H}^{\rm 2d}+\mathcal{H}^{\rm imp}$ with $V=t'^2/U$. To demonstrate this we numerically diagonalized this tight-binding model and calculated the Majorana polarisation and the energy corresponding to the lowest-energy eigenvalue, see the main text for more details. The results can be seen in Fig.~\ref{figs}, where these quantities are plotted as a function of $\mu$ and $V={t'}^2/U$. Note that we obtain a perfect agreement to Fig.~\ref{figs}.

\section{Analytical result for the chain Green's function}

We start with the momentum-space Hamiltonian for the two-dimensional superconducting substrate with Rashba spin-orbit coupling:
\begin{align}
\mathcal{H}^{\mathrm{2D}}_{\bs k} = - \xi \tau_z - 2\lambda \left[\sin k_x \sigma_y - \sin k_y \sigma_x  \right] \tau_z - \Delta \tau_x,
\end{align}
where $\xi \equiv \left[\mu + 2t(\cos k_x + \cos k_y) \right]$. In what follows we first calculate the unperturbed Green's function of the bulk using $\mathcal{G}_0(\omega, \bs{k}) = \left(\im \omega + \mathcal{H}^{\mathrm{2D}}_{\bs k} \right)^{-1}$. For the sake of simplicity we perform this calculation only at $\omega = 0$. Thus we get:
\begin{align}
\mathcal{G}_0(0, \bs{k}) =& -\big\{ \left[\mu + 2t(\cos k_x + \cos k_y) \right] \tau_z 
	\\&
	+ 2\lambda \left[\sin k_x \sigma_y - \sin k_y \sigma_x  \right] \tau_z + \Delta \tau_x \big\}^{-1}
\end{align}
This inversion can be performed using the identity
\begin{equation}
\bpm M & \Delta  \\ \Delta & -M\epm^{-1} \equiv \bpm M (M^2 + \Delta^2 )^{-1} & \Delta (M^2 + \Delta^2 )^{-1} \\ \Delta (M^2 + \Delta^2 )^{-1} & -M (M^2 + \Delta^2 )^{-1} \epm
\end{equation}
where we denoted 
\begin{align}
M \equiv \bpm \xi & -2 \im \lambda(\sin k_x - \im \sin k_y)\\ 2 \im \lambda(\sin k_x + \im \sin k_y) & \xi \epm
\end{align}
and $\Delta$ is diagonal. Finally, we get:
\begin{widetext}
\begin{align}
\mathcal{G}_0(0, \bs{k}) =& -\frac{1}{2\Delta} \bigg[ \frac{1}{(\Delta - \im\xi)^2+4\lambda^2(\sin^2 k_x + \sin^2 k_y)} \bpm M_- & \Delta_- \\\Delta_- & -M_-\epm
	\\&\qquad+ \frac{1}{(\Delta + \im \xi)^2+4\lambda^2(\sin^2 k_x + \sin^2 k_y)} \bpm M_+ & \Delta_+ \\ \Delta_+ & -M_+\epm  \bigg],
\end{align}
where we defined
\begin{align}
M_\pm &= 
	\bpm 
		\xi(\Delta \pm \im \xi) \mp 4\im \lambda^2 (\sin^2 k_x + \sin^2 k_y) & -2\im \Delta \lambda (\sin k_x - \im \sin k_y) \\
		2\im \Delta \lambda (\sin k_x + \im \sin k_y) &  \xi(\Delta \pm \im \xi) \mp 4\im \lambda^2 (\sin^2 k_x + \sin^2 k_y)
	\epm \textrm{ and}
\\
\Delta_\pm &= 
	\bpm 
		\Delta (\Delta \pm \im \xi)  &  \mp 2\Delta \lambda (\sin k_x - \im \sin k_y) \\
		\pm 2\Delta \lambda (\sin k_x + \im \sin k_y) &  \Delta (\Delta \pm \im \xi)
	\epm \,.
\end{align}

To calculate the line Green's function $\mathcal{G}_1(0, k_y)$ at zero energy we need to integrate the bulk Green's function  $\mathcal{G}_0(0, \bs{k})$ found above over $k_x$ from $-\pi$ to $\pi$. All the odd integrands in $k_x$ yield zero, and hence we are left with the integrals
\begin{align}
\label{eq:I0def}I_n^\pm = \int_{-\pi}^\pi \frac{dk_x}{2\pi} \frac{\cos nk_x}{(\Delta \pm \im  \xi)^2 + 4 \lambda^2 (\sin^2 k_x + \sin^2 k_y)}\,,
\end{align}
where $n\in\{0,1,2\}$.
We note that for each integral $I_n^+(\Delta) = I_n^-(-\Delta)$.

The integrals can be straightforwardly performed in the complex plane using $z = e^{\im k_x}$. We rewrite the denominators of the integrands as follows:
\begin{align}
(\Delta - \im \xi)^2 + 4 \lambda^2 (\sin^2 k_x + \sin^2 k_y) =-\frac{t^2+\lambda^2}{z^2} (z-z_1)(z-z_2)(z-z_3)(z-z_4)\,,
\end{align}
where 
\begin{align}
z_{1,2} = \frac{1}{2}\left(w_- \mp \sqrt{w_-^2-4} \right)\,,\quad z_{3,4} = \frac{1}{2}\left(w_+ \mp \sqrt{w_+^2-4} \right)\,,
\end{align}
and $w_\pm \equiv -A \pm \sqrt{A^2+B}$ with
\begin{align}
A \equiv \frac{t(\mu + 2t\cos k_y + \im \Delta)}{t^2+\lambda^2},\textrm{ and }B \equiv \frac{4\lambda^2(1+\sin^2k_y)-(\mu + 2t\cos k_y + \im \Delta)^2}{t^2+\lambda^2}.
\end{align}
A simple analysis of the roots then shows that:
\begin{align}
(1)\; |z_{2,3}| < 1 \;\text{and}\; |z_{1,4}| > 1,\;\text{if}\quad 2 \lambda \sqrt{1+\sin^2 k_y} + \mu + 2t \cos k_y \geqslant 0,\; 2 \lambda \sqrt{1+\sin^2 k_y} -(\mu + 2t \cos k_y) \geqslant 0 \\
(2)\; |z_{2,4}| < 1 \;\text{and}\; |z_{1,3}| > 1,\;\text{if}\quad 2 \lambda \sqrt{1+\sin^2 k_y} + \mu + 2t \cos k_y > 0,\; 2 \lambda \sqrt{1+\sin^2 k_y} -(\mu + 2t \cos k_y) < 0 \\
(3)\; |z_{1,3}| < 1 \;\text{and}\; |z_{2,4}| > 1,\;\text{if}\quad 2 \lambda \sqrt{1+\sin^2 k_y} + \mu + 2t \cos k_y < 0,\; 2 \lambda \sqrt{1+\sin^2 k_y} -(\mu + 2t \cos k_y) > 0 
\end{align}
This means that every integral in Eqs.~(\ref{eq:I0def}) should be computed for the three different cases described above. We present the calculations below:
\begin{align}
I_0^- =& -\frac{1}{t^2 + \lambda^2} \frac{1}{2\pi \im} \oint\limits_{|z|=1} \negthickspace dz\; \frac{z}{\prod_{i=1}^4 (z-z_i)}=
	-\frac{1}{t^2 + \lambda^2} 
	\begin{cases}
	\frac{z_2}{(z_2-z_1)(z_2-z_3)(z_2-z_4)} + \frac{z_3}{(z_3-z_1)(z_3-z_2)(z_3-z_4)}
	& \textrm{(1)}\vspace{0.1cm}\\
	\frac{z_2}{(z_2-z_1)(z_2-z_3)(z_2-z_4)} + \frac{z_4}{(z_4-z_1)(z_4-z_2)(z_4-z_3)}
	& \textrm{(2)}\vspace{0.1cm}\\
	\frac{z_1}{(z_1-z_2)(z_1-z_3)(z_1-z_4)} + \frac{z_3}{(z_3-z_1)(z_3-z_2)(z_3-z_4)}
	& \textrm{(3)}
	\end{cases}\\
\nonumber\phantom{space}\\
I_1^- =& -\frac{1}{t^2 + \lambda^2} \frac{1}{4\pi \im} \oint\limits_{|z|=1} \negthickspace dz\; \frac{z^2+1}{\prod_{i=1}^4 (z-z_i)}=
	-\frac{1}{2}\frac{1}{t^2 + \lambda^2} 
	\begin{cases}
	\frac{z_2^2+1}{(z_2-z_1)(z_2-z_3)(z_2-z_4)} + \frac{z_3^2+1}{(z_3-z_1)(z_3-z_2)(z_3-z_4)}
	& \textrm{(1)}\vspace{0.1cm}\\
	\frac{z_2^2+1}{(z_2-z_1)(z_2-z_3)(z_2-z_4)} + \frac{z_4^2+1}{(z_4-z_1)(z_4-z_2)(z_4-z_3)} 
	& \textrm{(2)}\vspace{0.1cm}\\
	\frac{z_1^2+1}{(z_1-z_2)(z_1-z_3)(z_1-z_4)} + \frac{z_3^2+1}{(z_3-z_1)(z_3-z_2)(z_3-z_4)}
	& \textrm{(3)}
	\end{cases}\\
\nonumber\phantom{space}\\
I_2^- =& -\frac{1}{t^2 + \lambda^2} \frac{1}{4\pi \im} \oint\limits_{|z|=1} \negthickspace dz\; \frac{z^4+1}{ z \prod_{i=1}^4 (z-z_i)} = 
	-\frac{1}{2}\frac{1}{t^2 + \lambda^2}
	\begin{cases}
	1 + \frac{z_2^4+1}{z_2(z_2-z_1)(z_2-z_3)(z_2-z_4)} + \frac{z_3^4+1}{z_3(z_3-z_1)(z_3-z_2)(z_3-z_4)}
	& \textrm{(1)}\vspace{0.1cm}\\
	1 + \frac{z_2^4+1}{z_2(z_2-z_1)(z_2-z_3)(z_2-z_4)} + \frac{z_4^4+1}{z_4(z_4-z_1)(z_4-z_2)(z_4-z_3)} 
	& \textrm{(2)}\vspace{0.1cm}\\
	1 + \frac{z_1^4+1}{z_1(z_1-z_2)(z_1-z_3)(z_1-z_4)} + \frac{z_3^4+1}{z_3(z_3-z_1)(z_3-z_2)(z_3-z_4)}
	& \textrm{(3)}
	\end{cases}
\end{align}
The residue at $z=0$ is calculated using the fact that $\Pi_{i=1}^4 z_i = 1$ (a consequence of Vieta's formula). The remaining integrals $I_0^+, I_1^+, I_2^+$ can be computed by substituting $\Delta \to -\Delta$ in the expressions above.

The remaining part of the calculation boils down to expressing all the integrals of the Green's function in terms of the three integrals discussed above.

Finally, we can write the integral of the Green's function as
\begin{align}
\mathcal{G}_1(0, k_y) = \int\limits_{-\pi}^{\pi} \frac{dk_x}{2\pi}\mathcal{G}_0(0, \bs{k}) = \frac{1}{2\Delta} \sum\limits_{\sigma = \pm} \bpm \tilde{M}_\sigma & \tilde{\Delta}_\sigma \\ \tilde{\Delta}_\sigma & -\tilde{M}_\sigma \epm\,,
\label{eq:G1ky}
\end{align}
with
\begin{align}
\nonumber \tilde{M}_\pm^{11} =& \tilde{M}_\pm^{22} = \left[ \mu \Delta \pm \im (\mu^2+4t^2-4\lambda^2) + 2t (\Delta \pm 2\im \mu) \cos k_y \pm 2\im(t^2 + \lambda^2) \cos 2k_y \right] I_0^\pm \\
&\qquad\qquad+ \left[2t\Delta \pm 4\im t  (\mu + 2t \cos k_y)  \right] I_1^\pm  \pm 2 \im \left(t^2 +\lambda^2 \right) I_2^\pm\,, \\
\tilde{M}_\pm^{12} =& \tilde{M}_\pm^{21} =-2 \Delta \lambda \sin k_y  \cdot I_0^\pm\,,
\end{align}
and
\begin{align}
\tilde{\Delta}_\pm^{11} =& \tilde{\Delta}_\pm^{22} = \left[ \Delta^2 \pm \im \Delta (\mu + 2t \cos k_y)\right] I_0^\pm \pm 2\im t\Delta \cdot I_1^\pm\,, \\
\tilde{\Delta}_\pm^{12} =& \tilde{\Delta}_\pm^{21} =\pm 2 \im\Delta \lambda \sin k_y \cdot I_0^\pm\,.
\end{align}
\end{widetext}

%


\end{document}